\documentclass[prb,twocolumn,showpacs]{revtex4}
\usepackage{amsmath,amsfonts,amsthm,graphicx}
\usepackage{graphicx}
\usepackage{graphics}
\usepackage{dcolumn}
\usepackage{bm}

\newcommand{\beq}{\begin{equation}}
\newcommand{\eeq}{\end{equation}}
\newcommand{\beqnar}{\begin{eqnarray}}
\newcommand{\eeqnar}{\end{eqnarray}}
\newcommand{\bfig}{\begin{figure}}
\newcommand{\efig}{\end{figure}}

\begin{document}
\title{Control over band structure and tunneling in Bilayer Graphene \\ induced by velocity engineering}

\author{Hosein Cheraghchi}
\email{cheraghchi@du.ac.ir}  \author{Fatemeh Adinehvand}
\affiliation{School of Physics, Damghan University, 36716-41167,
Damghan, Iran}
\date{\today}
\newbox\absbox

\begin{abstract}
The band structure and transport properties of massive Dirac
Fermions in bilayer graphene with velocity modulation in space are
investigated in presence of the previously created band gap. It
is pointed out that the velocity engineering is considered as a
factor to control the band gap of symmetry-broken bilayer
graphene. The band gap is direct and independent of velocity
value if velocity modulated in two layers is set up equally.
Otherwise, in the case of interlayer asymmetric velocity, not
only the band gap is indirect, but also the electron-hole
symmetry fails. This band gap is controllable by the ratio of the
velocity modulated in the upper layer to the velocity modulated
in the lower layer. In more detail, the shift of momentum from
the conduction band edge to the valence band edge can be
engineered by the gate bias and velocity ratio. A transfer matrix
method is also elaborated to calculate four-band coherent
conductance through a velocity barrier possibly subjected to a
gate bias. Electronic transport depends on the ratio of velocity
modulated inside the barrier to the one for surrounding regions.
As a result, a quantum version of total internal reflection is
observed for enough thick velocity barriers. Moreover, a
transport gap originating from the applied gate bias is
engineered by modulating velocity of the carriers in the upper
and lower layers.
\end{abstract}


\maketitle
\section{Introduction}
Charge carriers in monolayer graphene at low energies, near the
neutrality point, are described by Dirac fermions with a velocity
that is independent of wavelength~\cite{RMP}. This unique
property proposes an analogous between Dirac fermions and
electromagnetic or mechanical waves in optics and acoustics.
Furthermore, this brings several unusual electronic properties
such as anomalous integer~\cite{QHE} and fractional~\cite{kim}
quantum Hall effects, electronic focusing by means of a
rectangular potential barrier (Veselago lensing)~\cite{veselago},
Klein tunneling ~\cite{exp-klein1,exp-klein2} and minimal
conductivity~\cite{minimal}.
\begin{figure}
\centering
\includegraphics[width=8cm]{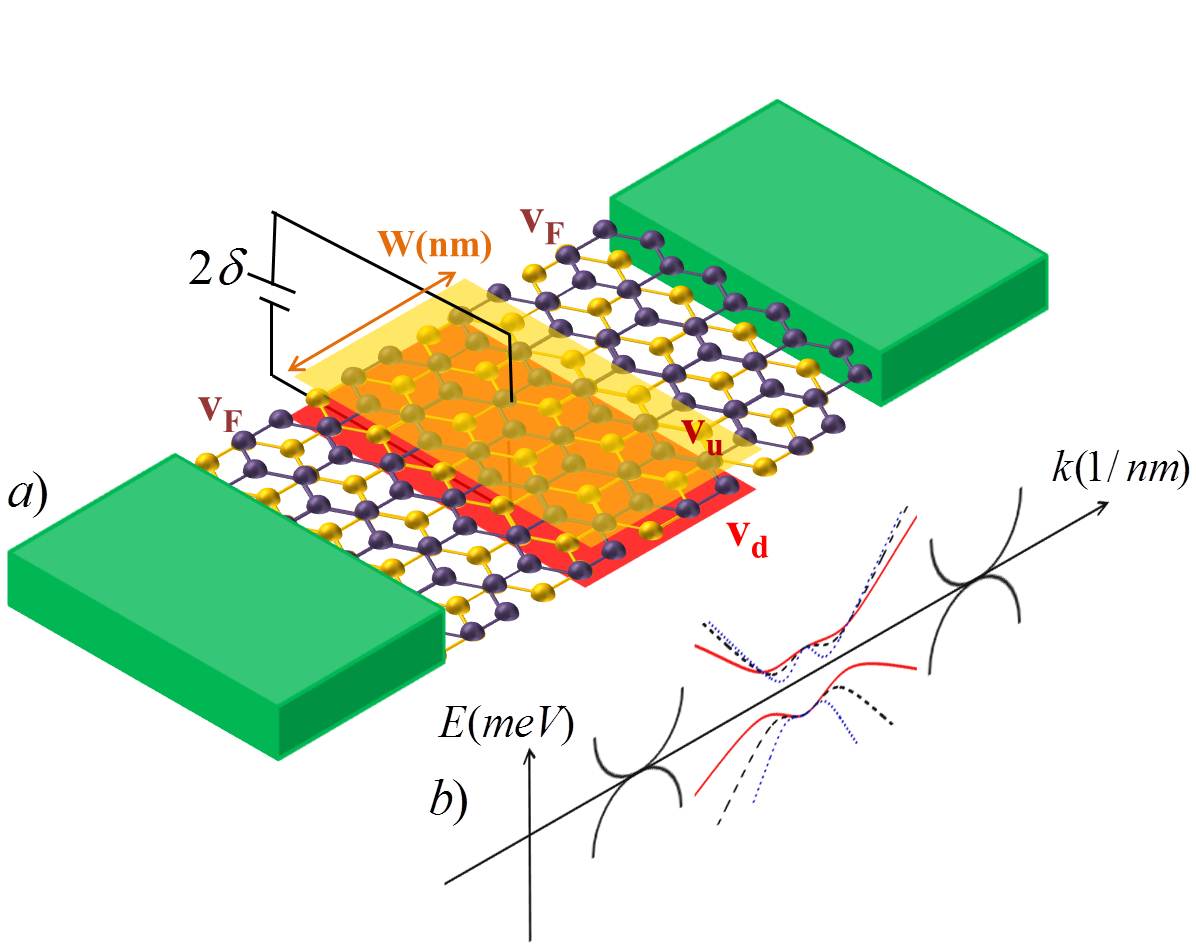}
\caption{ a) Schematic diagram of bilayer graphene junction in AB
stacking with velocity modulation in space. b) Energy band
structure of bilayer graphene for different regions with
different velocities. At the same time which vertically gate bias
$\delta$ is present, velocity may be experimentally modulated in
each layer of BLG. The ratio of velocity modulated in the upper
layer ($v_u$) to the lower layer ($v_d$) controls the feature of
the spectrum as well as tunneling through a velocity barrier.
}\label{schematic}
\end{figure}
Spatial modulation of wave velocity has been originally studied
in optics, acoustics and recently in photonic
crystals~\cite{photonics}. The idea can be also applied for Dirac
fermion waves by defining a velocity barrier as the region in
which the Fermi velocity differs from the one in the surrounding
background. In analogous with optics, some optical rules are
expected to be valid for massless Dirac fermion waves propagating
in monolayer graphene sheets~\cite{Concha}.

There are several ways to engineer the Fermi velocity ($v_F$) by
means of a control over the electron-electron interaction in
graphene. Enhancement in the electron-electron interaction
induces an increase in the Fermi velocity~\cite{olevano}.
Furthermore, an enhancement in $v_F$ which is logarithmic in the
carrier concentration $n$ has been established in experiments and
also described by the renormalization group theory~\cite{Elias}.
Modifications of curvature of graphene sheet~\cite{curvature},
periodic potentials~\cite{potential} and dielectric
screening~\cite{dielectric1,dielectric2} are some of propositions
for engineering $v_F$ via the electron-electron interaction. The
$v_F$ of graphene is inversely proportional to the dielectric
constant of the environment embedding graphene
sheet~\cite{hwang}. Structures with velocity modulation in space
can be also made by application of appropriate
doping~\cite{doping_velocity} or placing a grounded metal plane
as a screening plane close to graphene~\cite{raoux}. In the
presence of the screening planes, speed of carriers is smaller
than the speed at isolated graphene sheet. Recently, in a 2d
electron gas, an artificial graphene has been proposed by
modulating a periodic potential of honeycomb symmetry~\cite{AG}.
Electrons in artificial graphene sheets behave like massless
Dirac fermions with a tunable Fermi velocity.

The electronic properties of monolayer graphene sheets with
spatial modulation of the Fermi velocity have been investigated in
literature~\cite{Concha,raoux,JAP,Vasilopoulos,peres}. However,
the electronic properties in bilayer graphene (BLG) with an
interlayer asymmetric velocity have not been elucidated in detail
so far. There are numbers of different experiments in which a
controllable direct band gap is observed in gated bilayer graphene
~\cite{4band,gap-BG,gap-BG1,gap-BG2}. However, the amount of
current in the off-state still remains
high~\cite{gap-BG1,Martin,bilayer-review,hofmann}. This
off-current has been attributed to several sources
~\cite{bilayer-review}such as edge states ~\cite{Li}, the
presence of disorder~\cite{disorder-gap}, coexistence of massive
and massless Dirac fermions in twisted AA-stacking bilayer
graphene grown on SiC ~\cite{Bostwick,hofmann}. Strain is the
other known factor which controls the band gap in
BLG\cite{choi,verberck}.

In this work, we point out that the velocity modification in
symmetry-broken BLG, as an inevitable experimental factor, is
able to control the band gap. In the absence of the gate bias
$\delta=0$, symmetric or asymmetric velocity modulation in two
layers is not able to create a gap in the band structure.
However, the previously created gap $\delta\neq0$ can be
controlled by the ratio of modulated velocity in the upper layer
to the lower layer $\eta$. The band gap is direct if velocity of
itinerant quasi-particles in each layer is set up equally. This
gap is independent of velocity, while the momentum attributed to
the band gap is inversely proportional to the velocity. On the
other hand, the band gap is indirect for non-equal velocities
modulated in layers. In this case, the band structure and
subsequently the band gap are controlled by $\eta$. The Shift of
momentum from the conduction band edge to the valence band edge
depends on the gate bias and velocity ratio. Moreover, the
electron-hole symmetry fails when $\eta\neq1$. This kind of
control over the band structure which is induced by different
velocity modulation in each layer, opens up the possibility of new
device applications in nanoelectronics. More importantly, in a
BLG under application of gate bias, experiments have to be care
about the transition of direct to indirect band gap. This
transition can be induced by modification of velocity in layers
originating from several experimental requirements such as
coating a metallic gate electrode, changing carrier concentration
by using application of a gate voltage, strain and etc.

To manifest such a control over the gap, we develop a transfer
matrix approach to investigate transport properties through the
velocity barrier subjected to a gate bias in BLG. A schematic
diagram of the proposed system is presented in
Fig.\ref{schematic} which indicates simultaneous velocity and
electrostatic junction. The proposed method is based on a
four-band Hamiltonian for AB stacking~\cite{barbier,peeters}. As a
result, similar to monolayer
graphene~\cite{Concha,raoux,JAP,Vasilopoulos,peres}, a total
internal reflection occurs for Dirac fermion waves hitting on a
thick barrier at the angles of incidence greater than a critical
angle. Moreover, it is observed that the transport gap depends on
the velocity ratio $\eta$ at large gate bias. This gap is induced
by application of a symmetry breaking factor in the barrier
region.

We organize this paper as the following: In section II, we present
four-band Hamiltonian and a general formula for deriving the
spectrum in the presence of velocity modulation in addition to
vertically applied gate bias. Then in section III, we switch to
calculate transport properties though a velocity junction
possibly subjected to an external gate bias in generic form.
Finally, the last section includes the results.
\section{Hamiltonian and band structure in presence of interlayer asymmetry}
The four-band Hamiltonian of bilayer graphene close to the Dirac
point (i.e say the valley of $K$ point) for AB stacking is
described as the follow:
\beq  H= \begin{pmatrix}
-i \hbar v_{u} (\sigma . \nabla )^{\dag}+V_{u}I  & F \\
F & -i \hbar v_{d} (\sigma . \nabla )+V_{d}I
\end{pmatrix}\label{hamiltonian}
\eeq
where
$$ F=\begin{pmatrix}  t & 0 \\0 &0 \end{pmatrix}, -i \hbar v\sigma.\nabla =
\begin{pmatrix}0&\pi ^{\dag}\\\pi&0\end{pmatrix}$$ and $I$ is the
unit matrix. Here, $\pi=-i \hbar v(\partial _{x}-k_{y})$, $t=390
meV$ is the coupling energy between the layers. $V_{u}=V_0+\delta$
and $V_{d}=V_0-\delta$ describe an asymmetric factor which can be
applied by a vertically gate bias or doping. This interlayer
asymmetry emerges as a difference between on-site energies
belonging to each layer. Another interlayer asymmetry can be
induced by different modulation in the velocity of itinerant
quasi-particles in the upper and lower layers, $v_{u}=\xi_u v_F$
and $v_{d}=\xi_d v_F$ respectively. $v_F$ is the commonly Fermi
velocity used for graphene. $V_0$ is the gate voltage applied on
both layers setting up to zero. $2\delta$ is the potential
difference between the upper and lower layers induced by a gate
bias or doping. The eigen function of the above
Hamiltonian~\cite{barbier} is written as $ \Psi =
\begin{pmatrix} \psi_{A_2}^{u}&\psi_{B_2}^{u}&\psi
_{B_1}^{d}&\psi _{A_1}^{d} \end{pmatrix}^{\top}$. By solving the
eigenvalue equation of $ H\Psi= E\Psi$, band structure can be
calculated in the gapless case or in the presence of previously
applied gate bias.

\begin{figure}
\centering
\includegraphics[width=7cm]{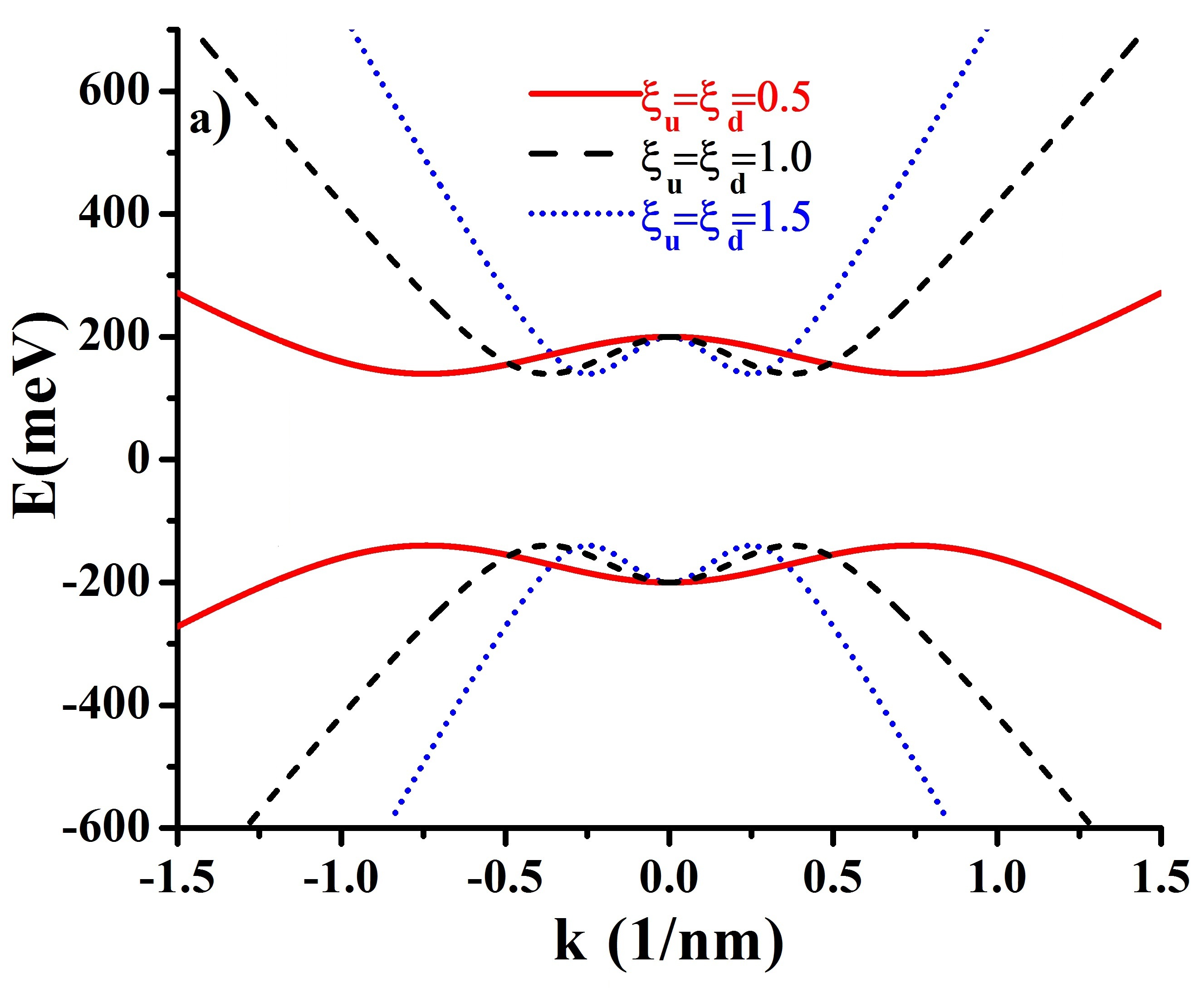}
\caption{Bulk band structure of bilayer graphene for the case of
the same velocity modulated in the upper and lower layers whenever
a band gap is previously created by application of a vertically
gate bias. Note that in this case, velocity in two layers are
equal to each other but possibly can be different from $v_F$. In
other word, $\xi=\xi_u=\xi_d$. }\label{band-gap-u=d}
\end{figure}

 At the same time which vertically gate bias is present,
velocity may be experimentally modulated in each layer of BLG.
\begin{figure}
\centering
\includegraphics[width=7cm]{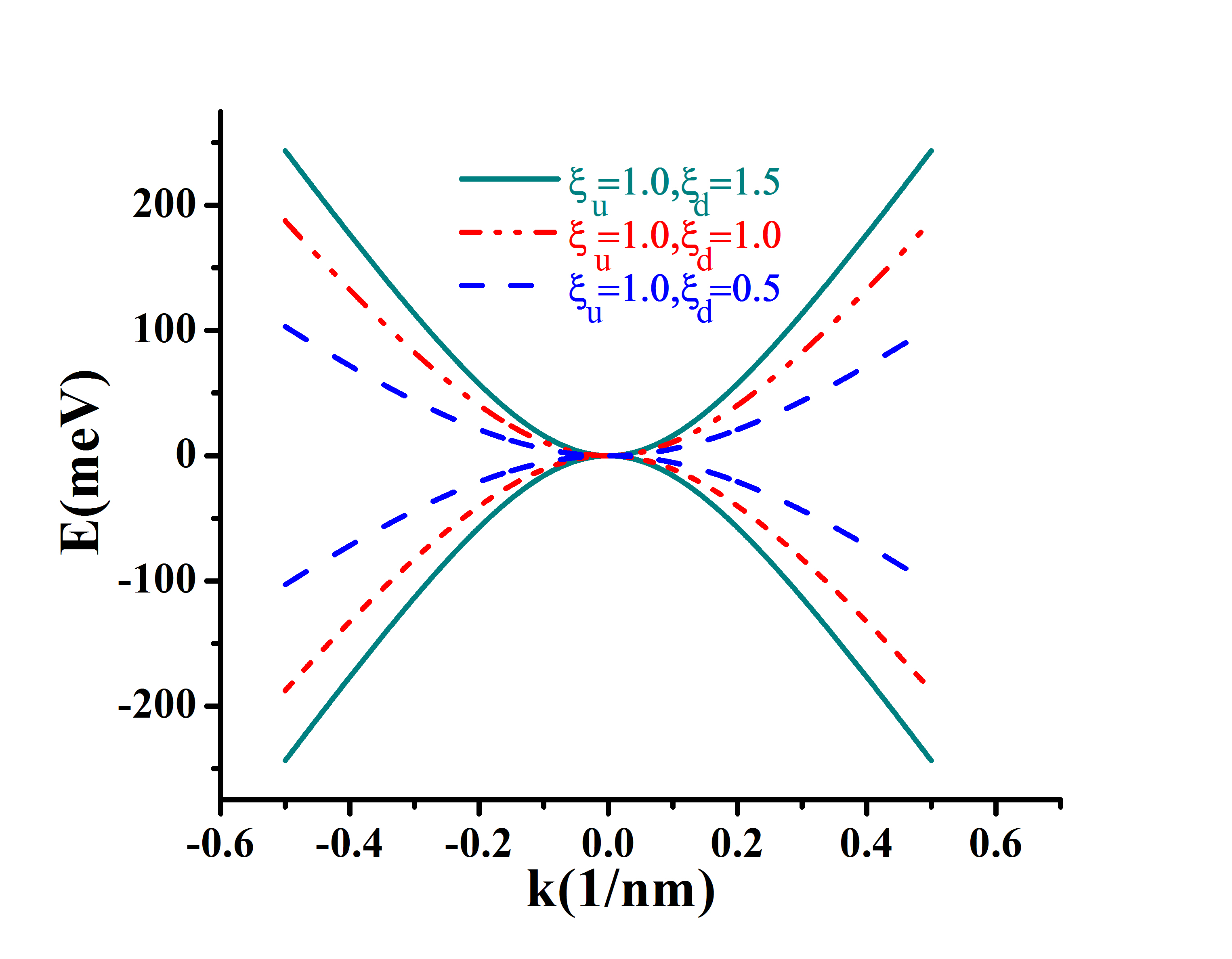}
\caption{ Bulk band structure of bilayer graphene for several
values of velocity modulated in the lower layer $\xi_d$ while
velocity in the upper layer ($\xi_{u}=1$) is fixed. Here,
$\xi=v/v_{F}$. }\label{band-nogap}
\end{figure}
In a gaped BLG, we will show that there is possibility for
engineering the previously created gap by using a velocity
modulation in each layer. In the presence of a gate bias
accompanied with an interlayer asymmetry in velocity, the BLG's
spectrum can be extracted from the following equation which
presents $k(E)$ .
 \begin{equation}
  \begin{aligned} &k(E)^{2} =[a\pm \sqrt{a^{2}-b}]/v_u^{2}\\
&a(E,\eta,\delta)=[\eta^{2}(E-\delta)^{2}+(E+\delta)^{2}]/2\\
&b(E,\eta,\delta)=\eta^2(E^{2}-\delta^{2})(E^{2}-\delta^{2}-t^{2})\\
&\eta=\xi_u / \xi_d  \label{spectrum}\end{aligned}
 \end{equation}
If the gate voltage is nonzero $V_0\neq0$, functions of $a$ and
$b$ in the above equation depend on $\varepsilon=E-V_0$ instead
of $E$. Based on the velocity ratio ($\eta$) and in the presence
of a gate bias ($2\delta$), we will indicate that BLG has two
different behaviors. For $\eta=1$, BLG behaves as a semiconductor
with direct band gap, while for $\eta\neq1$, it behaves as a
semiconductor with indirect band gap. In the case of $\delta=0$,
independent of $\eta$, there is no gap in BLG. Bulk band
structure calculated by the above equation for $\delta=0$ is shown
in Fig.(\ref{band-nogap}) and for $\delta \neq 0$ in
Fig.(\ref{band-gap-u=d}) and Fig.(\ref{band-gap-u}a).

To investigate the behavior of the energy gap $E_{gap}$, one can
simply derive the following conditions to emerge the extermum
points of $E(k)$. Based on Eq.(\ref{spectrum}), there are two
conditions to satisfy the extermum condition $\partial E/\partial
k=0$:

\beq \left\{ \begin{aligned} & b=0  & \forall  &\,\,\ k=0 \\
&b=a^2  & \forall &  \,\ k=\pm v_u^{-1} \sqrt{a}
\end{aligned} \right.\label{bandgap-conditions}\eeq

An immediate result from Eq.\ref{bandgap-conditions} is that the
energy gap $E_{gap}(\eta)=E_c(\eta)-E_v(\eta)$ depends on the
velocity ratio $\eta$, where $E_c$ and $E_v$ are the conduction
and valence band edges; however, the momentum attributed to the
conduction $k_c$ and valence $k_v$ band edges depend on both
variables of $\xi_u$ and $\xi_d$. The condition $b=0$ results in
four eigenvalues of Hamiltonian at the Dirac point calculated as
$E= \pm \delta$ and $\pm (t+\delta)$. These eigenvalues and
consequently the energy gap appeared at $k=0$ are independent of
the velocity ratio $\eta$. The condition $b=a^2$ leads to the
energy gap at the k-points derived by the following equation:
$k_{c/v}(\xi_u,\xi_d)=\pm v_u^{-1} \sqrt{a(E_{c/v},\eta,\delta)}$.

\subsection{Gapless band structure in presence of interlayer symmetric potential}

Let us first concentrate on the gapless case with no external gate
biasing ($\delta=0$) which conserves chiral symmetry. Based on
Eq.(\ref{spectrum}), the four band spectrum for a BLG with a
tunable velocity in each layer ($v_u \neq v_d$) can be derived as
the following~\cite{bilayer-review},

\beq \begin{aligned} &E =\pm\sqrt{\varphi(k)+(-1)^\epsilon
\sqrt{\varphi^2(k)-v_{u}^{2}v_{d}^{2}k^{4}}}\\
 &\varphi(k)=((v_{u}^{2}+v_{d}^{2})k^{2}+t^{2})/2\end{aligned}.\label{gapless-spectrum}
\eeq where $\epsilon=1$ and $2$ are attributed to the low and
high energy bands, respectively. In the case of $\epsilon=1$,
there is no band gap at the Fermi Dirac point ($k=0$). The whole
spectrum is robust against the exchange of $v_u$ by $v_d$. This
robustness can also be derived by exchanging of $\eta\rightarrow
1/\eta$ in Eq.(\ref{spectrum}). In this case, the only real
solutions for the extermum points derived by
Eq.(\ref{bandgap-conditions}), are $E=0$ and $\pm t$ which emerge
at $k=0$.

The chiral symmetry is conserved even though quasi-particles have
different velocities in each layer. In this case, modulation of
velocities in each layer just changes the effective mass of
quasi-particles. Fig.(\ref{band-nogap}) shows the energy bands of
BLG with different velocities in each layer. The band structure
is symmetric and behaves as a parabolic form. As a conclusion,
without any application of potential difference, only interlayer
asymmetry in velocity is not able to break the electron-hole
symmetry.
\subsection{Band structure in presence of interlayer asymmetry in potential but symmetry in velocity: Direct band gap}
In the presence of an interlayer asymmetric factor such as an
external gate bias ($\delta \neq 0$) and in the special case of
the same velocities setting up on each layer $v_u = v_d=v$
($\eta=1$), the four band spectrum is described
as~\cite{bilayer-review}:

\beq
E^{2}=(vk)^{2}+\delta^{2}+t^2/2+(-1)^{\epsilon}\sqrt{(vk)^{2}(4\delta^2+t^2)+t^4/4
+k^{2}}\eeq

As shown in Fig.(\ref{band-gap-u=d}), the low energy band
$\epsilon=1$ displays a {\it Mexican hat} shape. Despite turning
external gate bias on, the band structure still remains symmetric
giving rise the electron-hole symmetry. The functions of
$a(E,\delta)$, $b(E,\delta)$ defined in Eq.(\ref{spectrum}) are
{\it independent} of $\eta$. Therefore, the band gap is
independent of the velocity which is modulated in layers. The
requisite condition for deriving the band gap ($b=a^2$ in
Eq.(\ref{bandgap-conditions})) results in a symmetric solution
for the conduction and valence band edges,

\beq E_c=-E_v=\dfrac{t\delta}{\sqrt{4\delta^{2}+t^{2}}}\,\,\
\forall \,\,\,\ k\neq 0 \label{gap-eta=1}\eeq

So the band gap is written as $E_{gap}=2E_c$. At $k=0$, the gap
is fixed to the value $2\delta$. Because $a(E_v)=a(E_c)$, one can
conclude that the momentum of the conduction and valence band
edges emerge at the same point $k_c=k_v=k_{gap}$ from the center
of valley.

\beq
k_{gap}=\pm\dfrac{2\delta}{v_F}\sqrt{\frac{t^2+2\delta^2}{t^2+4\delta^2}}\frac{1}{\xi}
\label{k-gap-eta=1}\eeq

Consequently, the band gap is {\bf direct} and the momentum
attributed to the gap is inversely proportional to the velocity
$\xi$. For the limit of low external gate bias $\delta \ll t$,
the band gap tends to the gate bias $E_{gap}\rightarrow 2\delta$.
However, for large potential differences $\delta \gg t$, the band
gap tends to saturate at the interlayer hopping energy
$E_{gap}^{sat.}\rightarrow t$. For both limits, the momentum
attributed to the band gap behaves as $k_{gap}\propto 2\delta
/v$. For the case of slower velocity $\xi <1$, the effective mass
at the conduction and valence band edges is heavier than the
effective mass for the one with faster velocity $\xi
>1$.

\begin{figure}
\centering
\includegraphics[width=7cm]{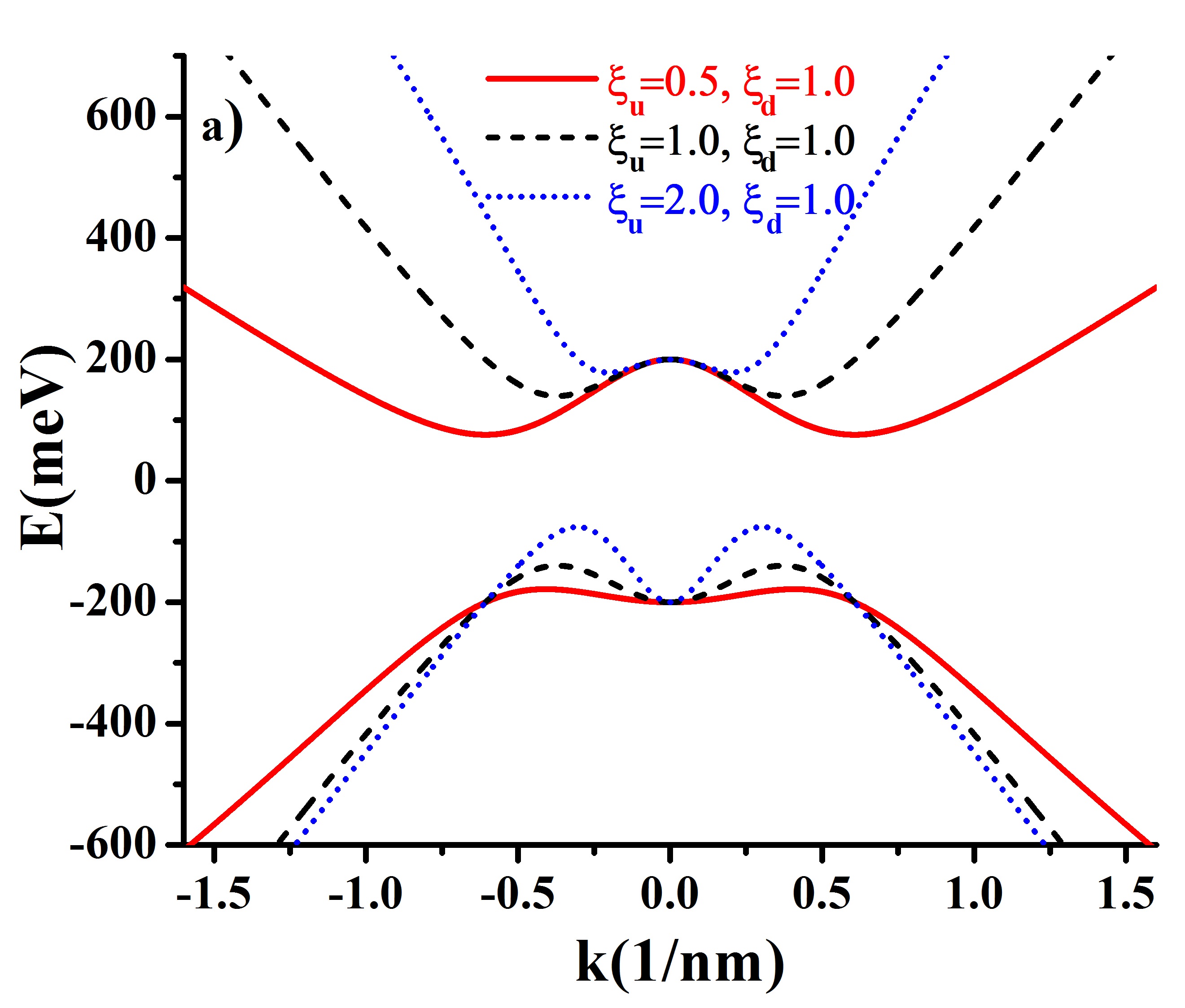}
\includegraphics[width=6cm]{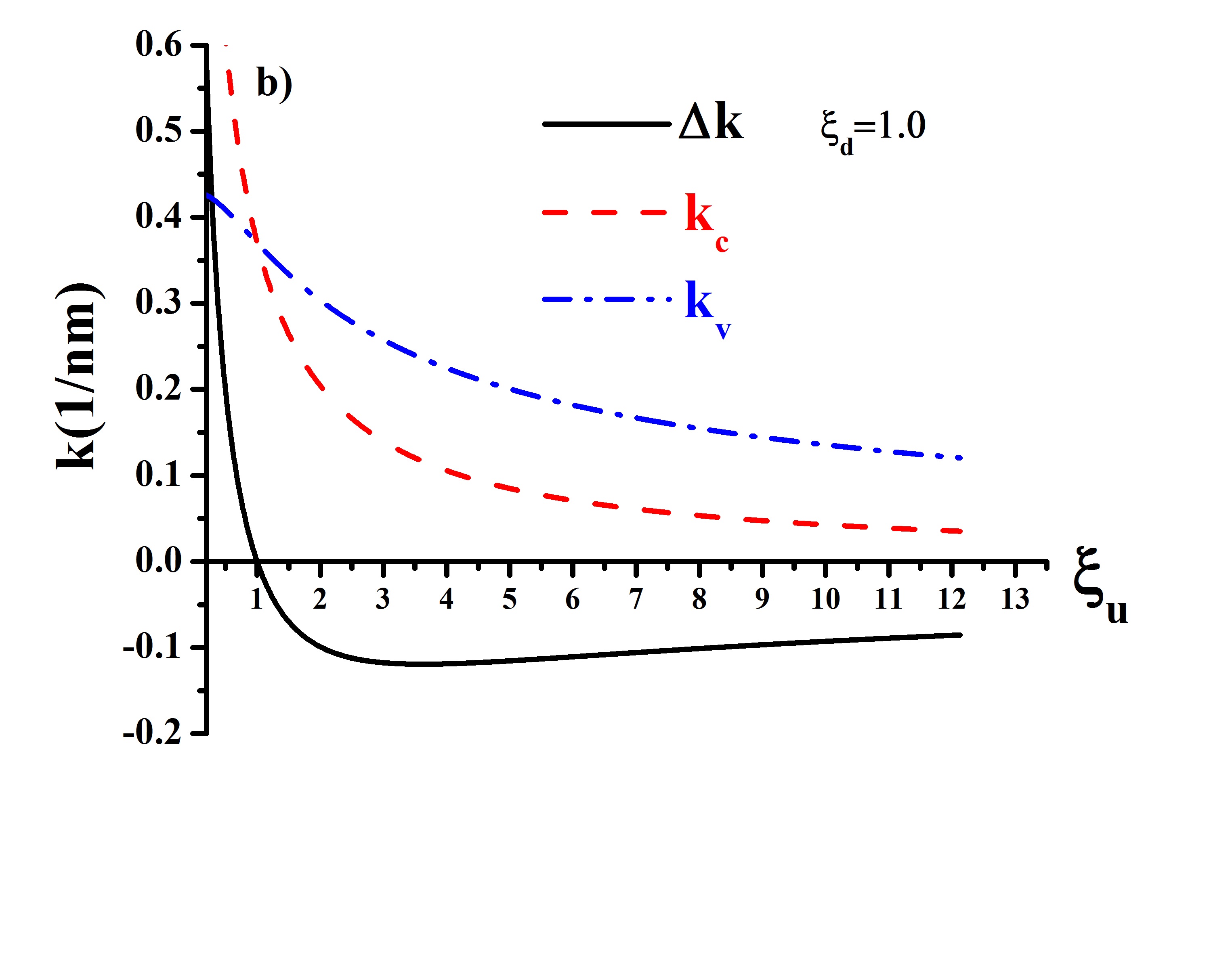}
\includegraphics[width=6cm]{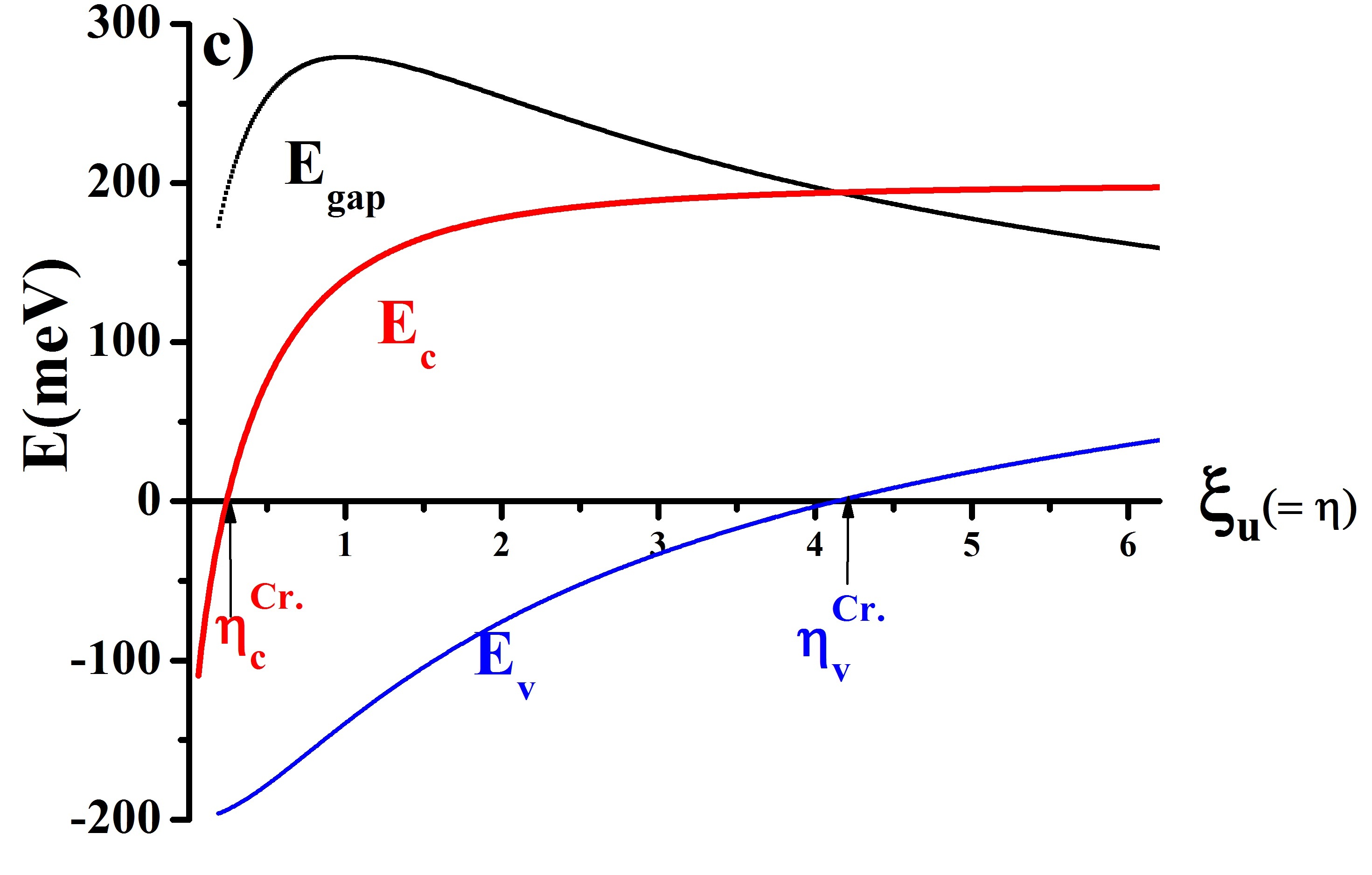}
\caption{a) Bulk band structure of bilayer graphene for an
interlayer asymmetry in velocity whenever a band gap is previously
created by application of a vertically gate bias. Note that
velocity modulation is $\xi_u$ for the upper layer and fixed for
the lower layer $\xi_d=1$. For more detail, the electrostatic
potential on the upper and lower layers are set $(+\delta)$ and
$(-\delta)$, respectively. Here the potential difference is set
$2 \delta=400 meV$. b) The momentum of the conduction $k_c$ and
valence $k_v$ band edges, and also the momentum shift from the
conduction band edge to the valence band edge $\Delta k=k_c-k_v$
as a function of $\xi_u$. c) The conduction $E_c$ and valence
$E_v$ band edges and also the energy gap $E_{gap}$ in terms of
$\xi_u$ (or here $\eta$). }\label{band-gap-u}
\end{figure}
\subsection{Band structure in presence of interlayer asymmetry in potential and velocity: Indirect band gap}
In this case, interlayer asymmetry is applied on both of
electrostatic potential and also velocity of itinerant
quasi-particles ($v_u\neq v_d$). In the case of $\eta \neq 1$ and
$\delta \neq 0$, there is an asymmetry between the conduction and
valence bands of the spectrum giving rise the electron-hole
asymmetry~\cite{ehasymm}. Consequently, the conduction and
valence band edges are appeared at asymmetric energy points
measuring from the band center $E=0$. As a result, the momentum
attributed to the conduction and valence band edges emerges at
different points, $k_c \neq k_v$. Therefore, the band gap is {\bf
indirect}. The shift of momentum from the conduction band edge to
the valence band edge ($\Delta k=k_c-k_v$) depends on velocity in
each layer.
\begin{figure}
\centering
\includegraphics[width=7cm]{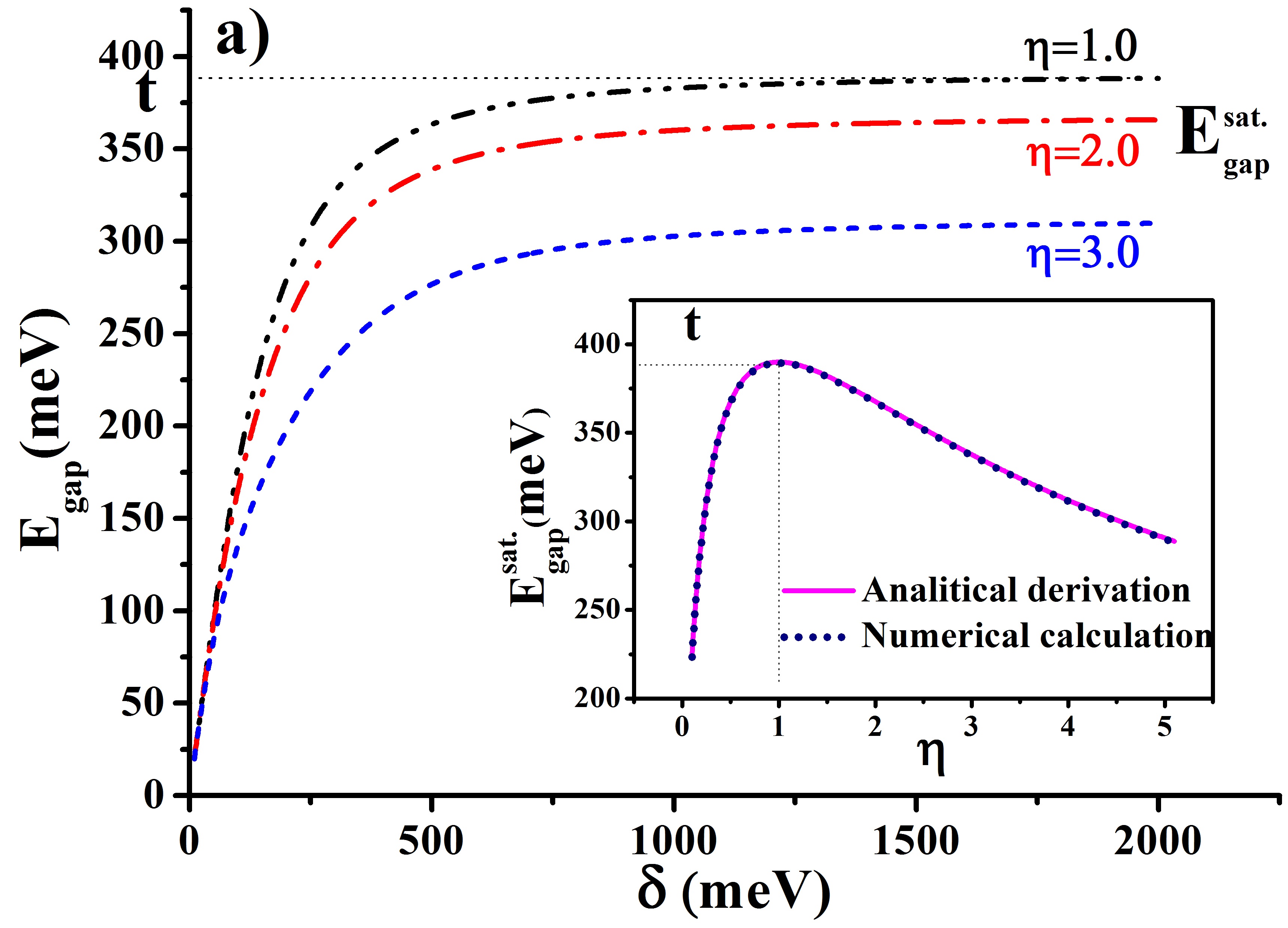}
\includegraphics[width=6cm]{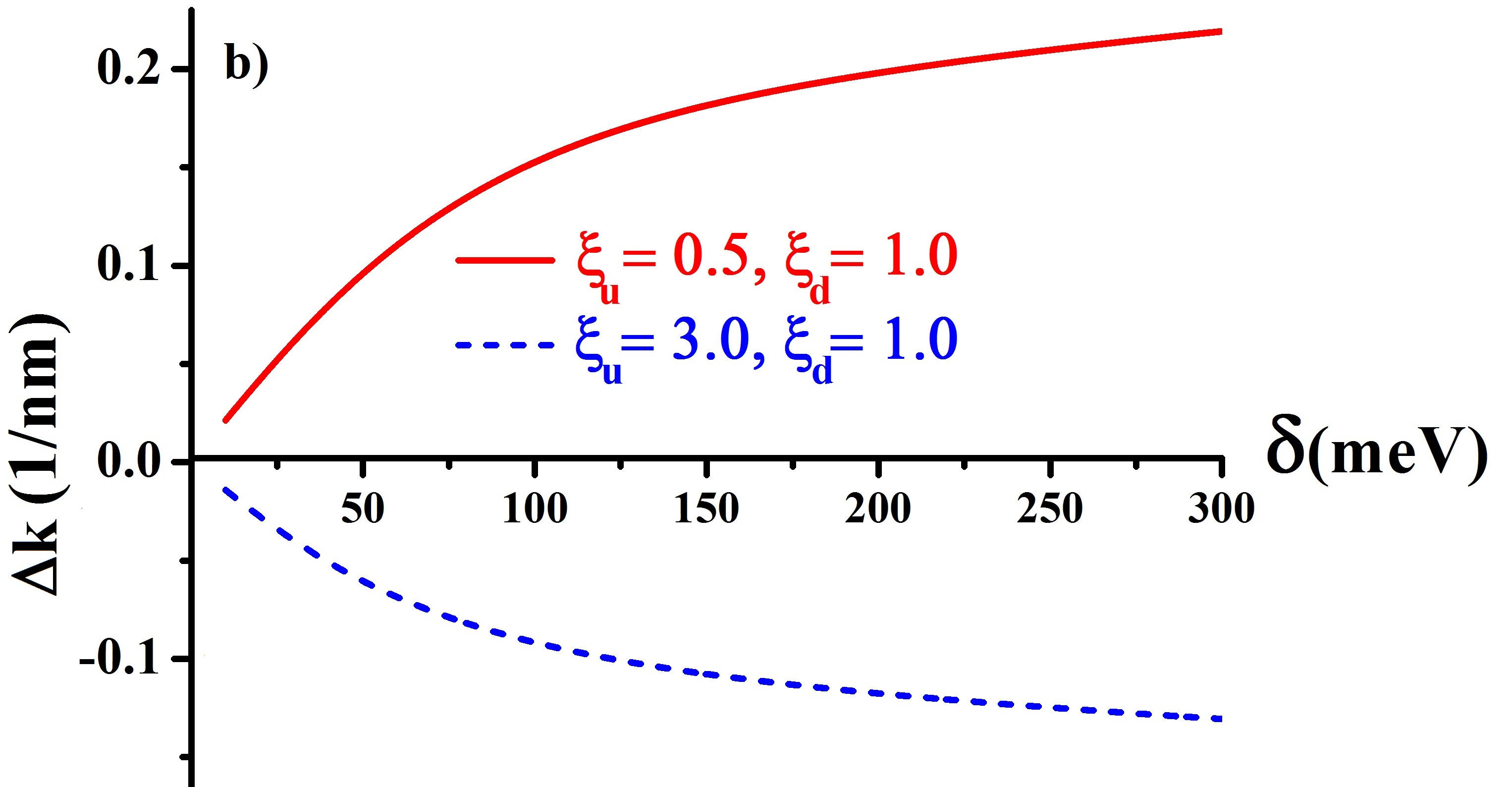}
\caption{a) The energy gap in terms of the potential difference
between the upper and lower layer for several velocity ratio. The
inset figure shows the saturated band gap in terms of the
velocity ratio for $\delta=4000 meV$. The numerical calculation
shown in the inset figure confirms the analytical derivation of
Eq.\ref{sat-gap}. b) The momentum shift from the conduction band
edge to the valence band edge in terms of the potential
difference between the upper and lower layer for several velocity
ratio. }\label{k-ph-delta}
\end{figure}
Although the band gap just depends on the velocity ratio $\eta$,
however, in a fixed velocity ratio, the whole feature of the
spectrum is sensitive to both values of velocity attributed to
the upper and lower layers. Let us set velocity of the lower
layer to be fixed as $\xi_d=1$ while $\xi_u$ is tunable. The
lower and upper layers are characterized by the electrostatic
potential of $-\delta$ and $\delta$, respectively.

The asymmetric band structure is represented in
Fig.(\ref{band-gap-u}a) for three values of velocity of the upper
layer $\xi_u$. Although the band structure is asymmetric, its
form preserves the '{\it Mexican hat}' shape. In the appendix
\ref{sec:appendix1}, we have provided a comparison between the
electron-hole asymmetry arising from the full Hamiltonian of BLG
and the dominant Hamiltonian which is considered in this work.

Fig.(\ref{band-gap-u}b) shows the momentum attributed to the
conduction $k_c$ and valence $k_v$ band edges and also their
momentum shift $\Delta k$ in terms of $\xi_u$. As it is obviously
observed, both of $k_c$ and $k_v$ decreases with $\xi_u$.
Moreover, their curves intersect each other at $\eta=1$ which
results in the direct band gap. However, for all values of $\eta
\neq 1$, the band gap is indirect. For $\xi_u<1$, the momentum
shift of $k_c$ away from the Dirac point is larger than the
momentum shift of $k_v$.

By finding roots of Eq.(\ref{bandgap-conditions}), the conduction
and valence band edges are computed in terms of system
parameters. Fig.(\ref{band-gap-u}c) indicates dependence of $E_c,
E_v$ and also $E_{gap}$ on the velocity ratio $\eta$. The curves
related to $E_c$ and $E_v$ never intersect each other. In all
ranges of $\eta$, $E_c>E_v$. So BLG always behaves as a
semiconductor, not metal nor semi-metal. The energy gap has a
maximum at $\eta=1$ in which the gap is direct. A sharp variation
of $E_{gap}$ with $\eta$ is seen for the range of $\eta<1$.
Parameters of $\eta^{cr.}_c$ and $\eta^{cr.}_v$ are those
critical velocity ratios in which $E_c$ or $E_v$ cross the band
center $E=0$. The curvature width of function $E_{gap}(\eta)$ is
measured by $\Delta \eta^{cr.}=\eta^{cr.}_v-\eta^{cr.}_c$. The
critical velocity ratio for the valence and conduction band edges
is derived as the following form: $
\eta^{cr.}_{v/c}=1+2(t/\delta)^2[1 \pm \sqrt{1+(\delta/t)^2}]$.
In both limits of $\delta \ll t$ and $ \delta \gg t$, the width of
the peak which emerges in $E_{gap}(\eta)$, tends to $\Delta
\eta^{cr.} \rightarrow 2t/\delta$. As a conclusion, for large
gate bias $\delta$, there is a sharp variation in the energy gap
as a function of the velocity ratio. In large velocity ratio
$\eta \rightarrow\infty$, the asymptotic solution of
Eq.(\ref{bandgap-conditions}) for the conduction band edge is
$E_c\rightarrow \delta$. In this limit, the momentum attributed
to the conduction band edge behaves as a power law with $v_u$;
$k_c \rightarrow 2\delta/v_u$. In the opposite limit of $\eta
\rightarrow 0$, the asymptotic solution for the valence band edge
is $E_v \rightarrow -\delta$. So, the momentum attributed to the
valence band edge tends to the constant; $k_v\rightarrow
2\delta/v_F$.

Although the energy gap increases with the external gate bias, as
it is shown in Fig.(\ref{k-ph-delta}a), the energy gap is
controllable by means of the velocity ratio in large $\delta$. In
fact, for $\delta \gg t$, the band gap saturates with the gate
voltage at the value which is proportional to the interlayer
coupling ($t$). In this limit, by applying the approximation of
($\mid E^2-\delta^2\mid\gg t^2$) in Eq.\ref{bandgap-conditions},
one can analytically derive that the saturated band gap at
$\delta \gg t$ behaves with the velocity ratio as the following
form;

\beq
E_{gap}^{sat.}(\eta)=\frac{2\sqrt{\eta}}{\eta+1}t.\label{sat-gap}
\eeq

In the special case of $\eta=1$, the band gap saturates at
$E_{gap}^{sat.}(\eta=1)\rightarrow t$.  As shown in the inset
Fig.(\ref{k-ph-delta}a), numerical calculations completely
confirm this analytical derivation. The momentum shift $\Delta
k$, which measures how much the gap is indirect, can be
manipulated by using the gate bias. Fig.(\ref{k-ph-delta}b)
represents the momentum shift from $k_c$ to $k_v$ in respect of
the gate bias for several values of $\xi_u$. This momentum shift
from $k_c$ to $k_v$ increases with the gate bias. If we transform
the velocity ratio as $\eta \rightarrow 1/\eta$, in the spectrum
feature, the conduction band will be exchanged with the valence
band. Furthermore, based on Eq.(\ref{bandgap-conditions}), the
band gap is robust against transformation of $\eta \rightarrow
1/\eta$.

In addition to the direct measurements of the spectrum, the
dependence of the energy gap on the velocity ratio can be
manifested in transport properties through a velocity junction.

\section{Transport properties across non-uniform potential and velocity junctions}
Let us consider a BLG sheet in which the velocity of itinerant
quasi-particles in the upper and lower layers varies in space;
representing as $v_u(\overrightarrow{r})$ and
$v_d(\overrightarrow{r})$. We assume that variation of velocity
is smooth on the scale of the lattice constant. In this section,
we outline the approach used to investigate transport properties
through a barrier of velocity and potential.

 \subsection{Current Density Operator}
First, by using the continuity equation, we derive the current
density operator. The continuity equation is as the following,
\beq  \nabla .j= - \partial_{t}\rho \label{continuity equation}
\eeq where  $ \rho =\Psi^{\dagger}\Psi $ is the charge and $ j $
is the current density operator. By using the Schroedinger
equation, divergence of the current density operator is written as
\beq
 \nabla .j = [(H\Psi)^{\dagger}\Psi -\Psi^{\dagger} (H\Psi)]/ i\hbar
\eeq

By substitution of $H$ from Eq.\ref{hamiltonian} and two
component spinor as $ \Psi =
\begin{pmatrix} \psi_{u}  \\ \psi_{d} \end{pmatrix}$ in the above
equation, we have

\beq
\begin{aligned}
i \hbar \nabla .j =-\begin{pmatrix}(-i \hbar v_{u}(\sigma
.\nabla)^{\dagger }+\delta) \psi_{u} +F\psi_ {d}\\ F\psi_{u}+(-i
\hbar v_{d} (\sigma . \nabla)-\delta)\psi_{d}
\end{pmatrix}^{\dagger}
\begin{pmatrix}\psi _{u }\\ \psi_{d } \end{pmatrix}\\+\begin{pmatrix} \psi _{u}^{\dagger}& \psi_{d }^{\dagger}\end{pmatrix}
\begin{pmatrix}(-i \hbar v_{u}(\sigma .\nabla)^{\dagger }+\delta)\psi _{u}+F\psi _ {d}\\ F\psi_{u}+(-i \hbar v_{d} (\sigma .\nabla)-\delta)\psi _{d} \end{pmatrix}
\end{aligned}
\eeq

After simplification, it is derived that interestingly, the
current density operator is independent of the gate bias $\delta$
and also the hopping matrix $F$.
 \beq
\begin{aligned}
 \nabla . j =  \left[ v_{u}\nabla.(\psi_{u}^{\dagger}\sigma^{\dagger } \psi_{u}) +
v_{d}\nabla .( \psi_{d}^{\dagger}\sigma \psi_{d})\right]
\end{aligned}
\eeq

Therefore, current density operator for a BLG sheet is presented
as,

 \beq j=\begin{pmatrix} \psi_{u}\\
\psi_{d}\end{pmatrix}^{\dagger}
\begin{pmatrix} v_{u}\sigma^{\dagger}&0\\ 0& v_{d}\sigma \end{pmatrix}
\begin{pmatrix} \psi_{u}\\ \psi_{d}\end{pmatrix}.
\eeq

Finally, the current density in the i'th region can be written in
the following compact form.

\beq j_{i}=\Phi_{i}^{\dagger} \Sigma \Phi _{i}
\label{current_density}\eeq where the auxiliary spinor is defined
as $\Phi_{i}=\widetilde{v}_{i} \Psi_{i}$ and

$$ \Sigma =\begin{pmatrix}\sigma^{\dagger } &0\\0&\sigma
\end{pmatrix},\,\,\,\,\widetilde{v}_{i}=\begin{pmatrix}
\sqrt{v^{i}_{u}}&0\\0&\sqrt{v^{i}_{d}} \end{pmatrix} $$.

\subsection{Transfer Matrix Method}
We assume a plane wave solution for the four-band Hamiltonian. So
the wave function in each region with a constant potential is
written as the following matrix product, $ \Psi(x) =P(x)*A$, where
$P(x)$ and $A$ are the plane wave and coefficient matrices,
respectively. Detail of matrices $P$ and $A$ are accessible in
appendix \ref{sec:appendix2} and also
Refs.(\onlinecite{cheraghchi-polarization,barbier}). The local
current density in terms of matrices $P(x)$ and $A$ in each
region reads as the following form,
 \beq
j_{i}=A_{i}^{\dagger}P_{i}^{\dagger}\widetilde{v}_{i}^{\dagger}\Sigma
\widetilde{v}_{i}P_{i} A_{i}
 \label{current_operator}\eeq
where the auxiliary spinor in Eq.\ref{current_density} has been
replaced by $\Phi_{i}=\widetilde{v}_{i}P_{i}A_{i}$. The continuity
equation of $\overrightarrow{\nabla} .
\overrightarrow{j}(\overrightarrow{r})=0$ leads to the boundary
matching condition at interfaces of a junction. On the other
word, conservation of the current density results in the
continuity of the auxiliary spinor $\Phi_{i}$ on the boundaries
of the barrier junction.
$$\Phi_1=\Phi_2  \Longrightarrow \widetilde{v}_{2}\Psi_2=\widetilde{v}_{1} \Psi_1 $$
Referring to the schematic cartoon shown in Fig.\ref{schematic},
we consider a simultaneous barrier of velocity,

\beq v(x)= \left\{ \begin{array}{c} v_{u}=v_d=v_F
\,\,\,\,\,\,\,\,\,\,\, {\rm {\bf I}:x<0,  {\bf III}:x>w}
\\ \\ v_u, v_d
\,\,\,\,\,\,\,\,\,\,\,\,\,\,\,\,\,\,\,\,\,\,\ {\rm {\bf II}:0<x<w
}\end{array} \right. \eeq

and electrostatic potential. At the same time, the barrier can be
subjected to a gate bias. \beq V(x)= \left\{
\begin{array}{c} V_u=V_d=V_0 \,\,\,\,\,\,\,\,\,\,\,{\rm {\bf I}:x<0,  {\bf III}:x>w}
\\ \\ \left\{ \begin{array}{c} V_u=V^{'}_0+\delta/2 \\ V_d=V^{'}_0-\delta/2 \end{array}\right.
\,\,\,\,\,\,\,\,\,\,\,\,\,\,\,\,\,\,\,\,\,\,\ {\rm {\bf II}:0<x<w
}
\end{array} \right.
 \eeq

By applying continuity of the auxiliary spinor on the boundaries
of the barrier, one can connect the coefficient matrix related to
the last region $A_3$ to the coefficient matrix for the first
region $A_1$.

\beq
\begin{array}{c}
A_{1} =M A_{3}\\ \\
M=P_{1}^{-1}(0)\tilde{v}_{1}^{-1}\tilde{v}_{2}P_{2}(0)P_{2}^{-1}(w)\tilde{v}_{2}^{-1}\tilde{v}_{3}P_{3}(w)
\label{continiuty}
\end{array}
\eeq where $M$ is the transfer matrix. We assume that the energy
range of incidence particles in the first region is limited to
the range of $0<\varepsilon_1<t$~\cite{cheraghchi-polarization}.
Consequently, the wave numbers $\alpha_+^{(1)}$ and
$\alpha_+^{(3)}$ which are defined in the
appendix.\ref{sec:appendix2}, are real while $\alpha_-^{(1)}$ and
$\alpha_-^{(3)}$ are imaginary. In this range of energy,
coefficient matrices in the first and third regions are proposed
as the following form.
\begin{equation*}
A_{1}=\begin{pmatrix}1&r&0&e_{g}
\end{pmatrix}^{\top} ,A_{3}=\begin{pmatrix}
t&0&e_{d}&0\end{pmatrix}^{\top}
\end{equation*}
For the first region, $e_g$ is the coefficient of growing
evanescent state and $r$ is the coefficient of reflection. In the
last region, $t$ is the transmission coefficient and $e_d$ is the
coefficient of decaying evanescent state. By rearrangement of
Eq.~\ref{continiuty}, the coefficient of transmission is derived
as a function of the transfer matrix elements as the following;

 \beq
t=[M_{11}-M_{13}M_{31}/M_{33}]^{-1}.\eeq
The transmission probability of particles through a barrier is
defined as the ratio of out-flowing current to in-flowing
current.
 \beq T=\dfrac{J_{out}^{3}}{J_{in}^{1}} \eeq
where $ J_{out}^{3} $ is the out-flowing current in the last
region and $ J_{in}^{1} $ is the in-flowing current incidence
from the first region. By using Eq.\ref{current_operator}, the
transmission probability can be represented as the following
form. \beq T=\dfrac{\begin{pmatrix}t&0&0&0
\end{pmatrix}
P_{3}^{\dagger}\tilde{v}_{3}^{\dagger}\Sigma \tilde{v}_{3}P_{3}
\begin{pmatrix}
t\\0\\0\\0 \end{pmatrix}}
 {\begin{pmatrix}
 1&0&0&0
\end{pmatrix}
P_{1}^{\dagger}\tilde{v}_{1}^{\dagger}\Sigma \tilde{v}_{1}P_{1}
\begin{pmatrix}
1\\0\\0\\0
\end{pmatrix}}
\eeq

The conductance is calculated by using Landauer formalism in the
linear regime. Transport is coherent and is calculated at zero
temperature. Conductance is proportional to angularly averaged
transmission projected along the current direction.

\begin{equation}
G=2G_{0}\int _{0}^{\pi/2} T(E,\varphi ) \cos(\varphi) d\varphi
\end{equation}
where $G_{0}=e^{2}mvw/\hbar^{2}$.
\begin{figure}
\centering
\includegraphics[width=6cm]{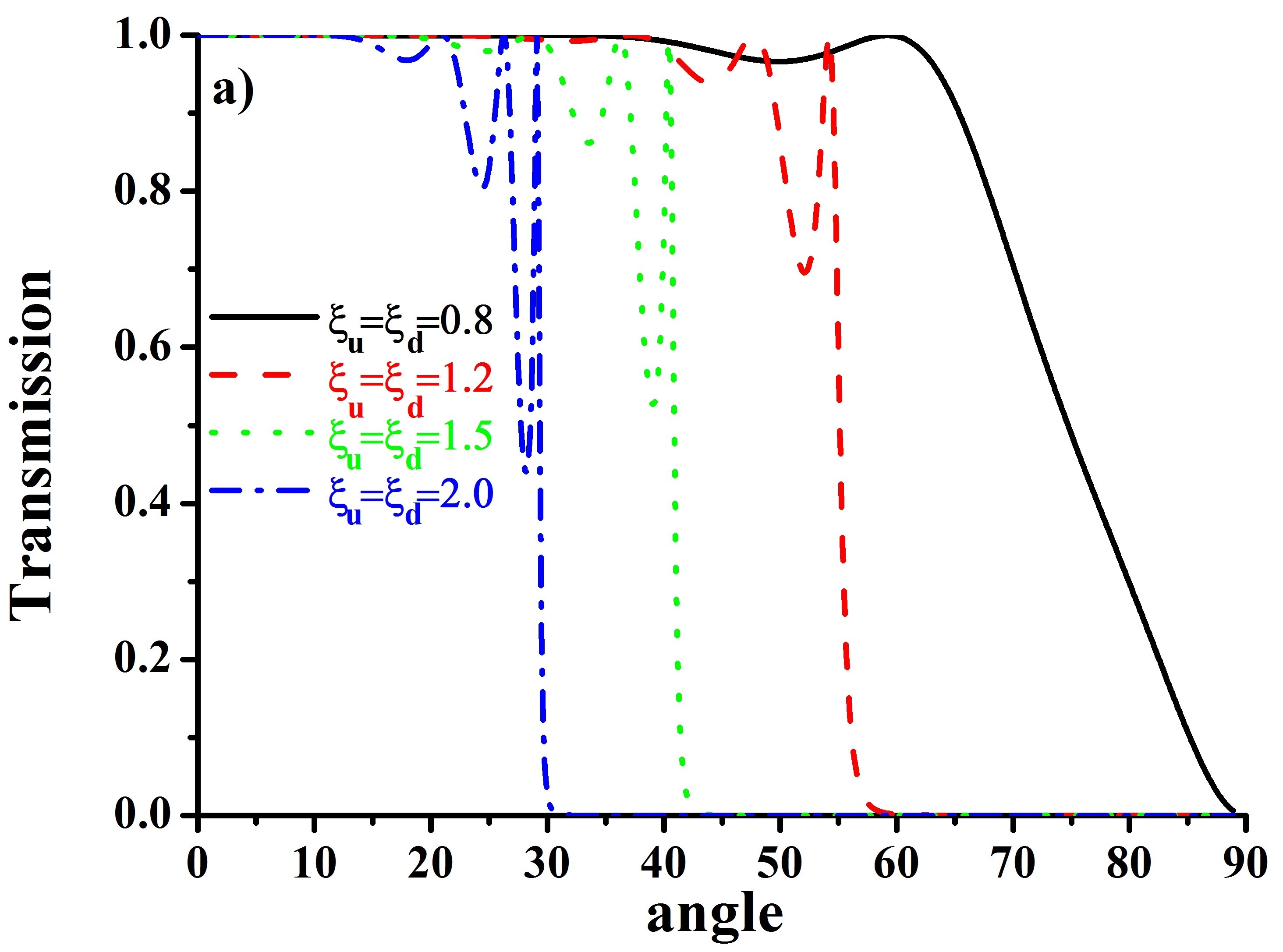}
\includegraphics[width=6cm]{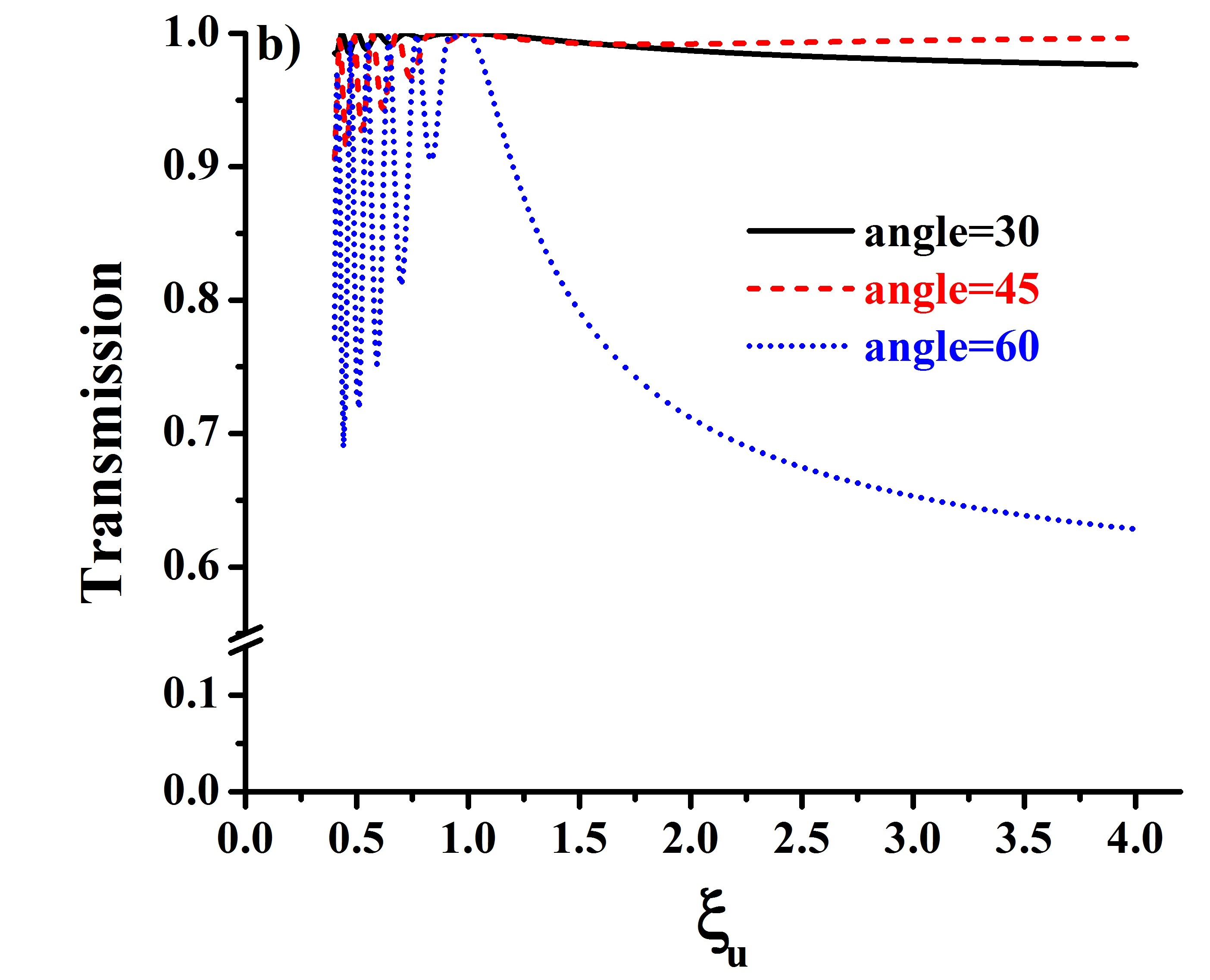}
\includegraphics[width=6cm]{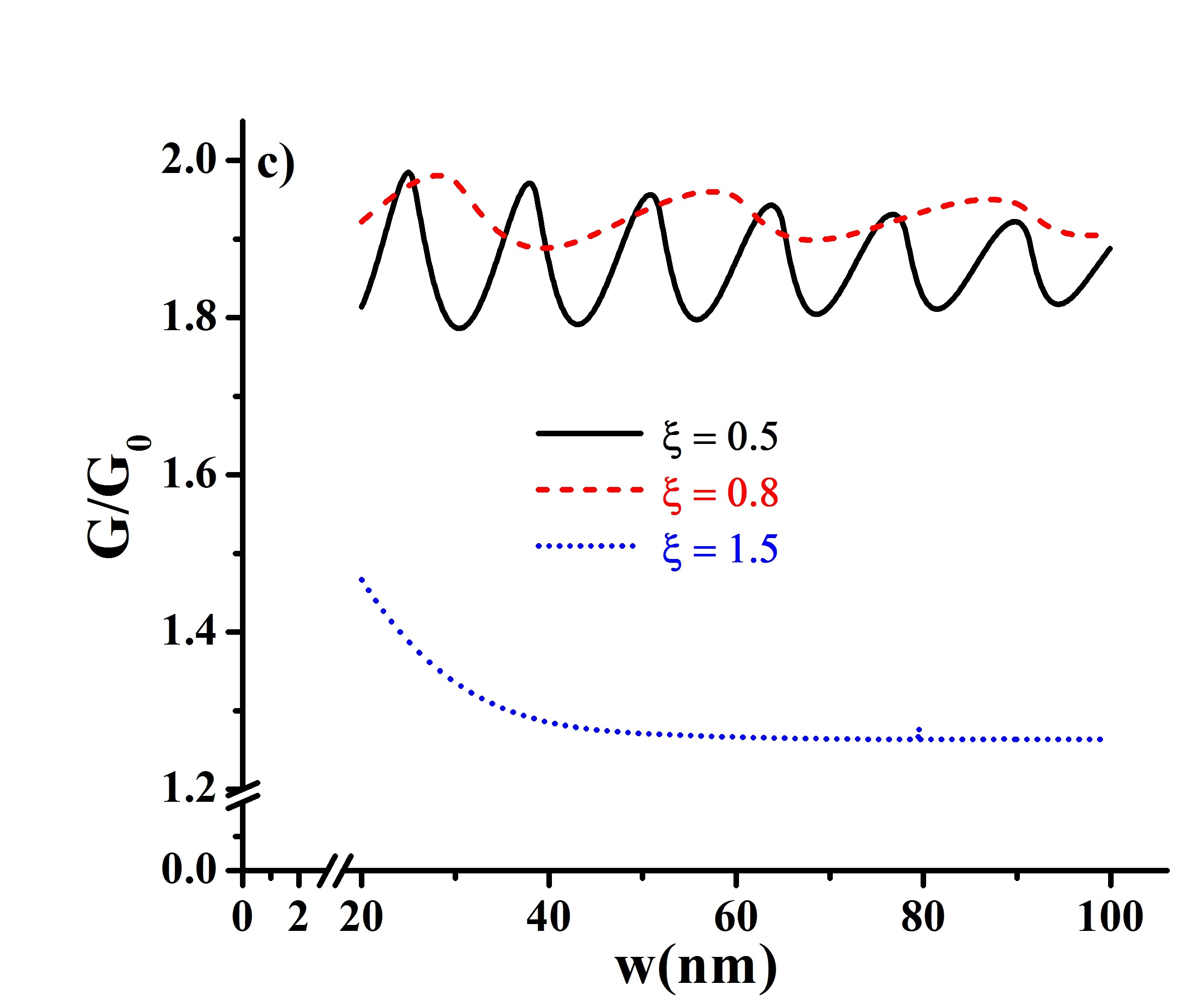}
\caption{a) Transmission probability as a function of incidence
angle $\theta_1$ for the case of the same velocity modulated in
both layers. The velocity ratio $\xi_u=\xi_d=v_2/v_F$ is set to
$0.8, 1.2, 1.5, 2.0$. Estimated critical angle for the velocity
ratio $1.2$ and $1.5$ is about $\theta_{cr.}\approx 56$ and $
42$. b) Transmission probability in terms of the velocity ratio
for several incidence angles. we consider a thick velocity barrier
 with the width $w=100 nm$ for parts (a) and (b). c) Conductance as a
function of the width for the velocity ratio equal to $0.5, 0.8,
1.5$. For all parts, the energy value is set to $E=10
meV$}\label{T_theta_xi}
\end{figure}
\subsection{Transport across a single velocity barrier}
The behavior of a beam produced by Dirac fermions whenever hit on
the barrier region, is similar to the behavior of an optical beam
passing through dielectric materials. In the subsequent sections,
we will show that a quantum mechanical version of well-known laws
in geometrical optics can be also applied on the propagation of
Dirac fermions in BLG.

{\bf Case i}: Let us consider tunneling through a single
velocity-induced sharp barrier. For a pure velocity barrier, type
of quasi-particles inside and outside of the barrier is the same
for all ranges of energy. For normal incidence $\theta_1=0$ and
in absence of any gate bias, transmission coefficient for a
velocity barrier with unity velocity ratio $\eta=1$ inside and
outside of the barrier can be analytically calculated as

\begin{equation}
 t=e^{i\alpha_{1}w}[\cos(\alpha _{2}w )-iS\sin(\alpha _{2}w) ]^{-1}
 \end{equation}
 where $$ S=\frac{1}{2}(\dfrac{\omega_{1}\alpha_{2}} {\omega_{2}\alpha_{1}}+\dfrac{\omega_{2}\alpha_{1}}{\omega_{1}\alpha_{2}}) $$
and
$\alpha_{1}=[\omega_{1}^{2}+\omega_{1}t/v_{1}]^{1/2},\alpha_{2}=[\omega_{2}^{2}+\omega_{2}t/v_{2}]^{1/2}$
are the wave vectors along the x-axis direction outside and inside
the velocity barrier, respectively. Here, scaled energy in each
region is defined as $\omega _{1}=E/ v_{1}$ and $\omega _{2}=E/
v_{2}$. Replacing defined parameters in $S$, results in $S=1$.
Therefore, transmission probability is derived as the following
form,

\begin{equation}
 T=\mid t \mid^2=\dfrac{1}{[\cos^{2}
 (\alpha_{2}w)+\sin^{2}(\alpha_{2}w)]}=1.
 \end{equation}
As a result, independence of all barrier parameters, transmission
at the normal incidence is always perfect. This behavior is
similar to what we expect from the standard Klein tunneling. This
transparency at the normal incidence will be demonstrated
numerically in Fig.\ref{T_theta_xi}. At arbitrary incidence
angle, the wave vectors along the
 x-axis direction in the regions I and II can be represented as the
 following.

 \begin{equation}
\alpha_1=\sqrt{\dfrac{1}{v_{1}^{2}}(E^{2}+tE)-k_{y}^{2}} ,\,\,\,\,
\alpha_{2}=\sqrt{\dfrac{1}{v_{2}^{2}}(E^{2}+tE)-k_{y}^{2}}
\end{equation}

Suppose that the velocity outside the barrier $v_1$ is set to be
$v_F$. Conservation of the energy $E$ and the component $k_y$ of
the wave vector across the barrier leads to the following compact
form for the wave vector inside the velocity barrier.
\begin{equation}
\alpha_{2}=k\sqrt{\dfrac{1}{\xi^{2}}-\sin^{2}\theta_{1}}
\label{wave_vector_barrier}
\end{equation}
where $\xi=\xi_u=\xi_d=v_2/v_F$ is the velocity of quasi-particles
inside the barrier scaled by $v_F$. $\theta_1 $ is the incidence
angle of quasi-particles which hit on the barrier from the region
I. For the range of $\xi>1$, a look at
Eq.\ref{wave_vector_barrier} obviously demonstrates that there
are some evanecsent modes in the barrier region (in which
$\alpha_2$ is imaginary) if only the incidence angle $\theta_1$
is greater than a critical angle which is defined as,

\begin{equation} \theta_{cr.}=\arcsin
(1/\xi).\end{equation}

In analogous with optics, the total internal reflection (TIR)
emerges when a Dirac fermion wave hits from a denser medium
(region I) on a rarer medium (the barrier region II). This
behavior is interpreted as $\xi>1$ in our studied
system~\cite{Concha,raoux,JAP,Vasilopoulos,peres}. To demonstrate
such a critical angle in BLG, we plot transmission probability
$T(\theta_1)$ as a function of the incidence angle in
Fig.\ref{T_theta_xi} for several values of velocity. For $\xi>1$
and $\theta_1>\theta_{cr.}$, transmission is negligible for enough
thick barriers. We have also checked that variation of
transmission around the critical TIR angle is more sharp for the
multiple structure of velocity barriers in compared with the
single velocity barrier. Furthermore, as indicated in
Fig.(\ref{T_theta_xi}b), transmission probability shows a sharp
change in behavior at $\xi=1$. In the case of $\eta\neq1$, the
larger velocity modulated in the upper or lower layer, the
smaller critical angle emerges. The critical angle just depends
on $\xi$. So this property is more appropriate for designing a
waveguide based on the BLG substrates\cite{JAP}.

As a conclusion for Eq.\ref{wave_vector_barrier}, for the range of
$\xi<1$, the wave vector inside the barrier $\alpha_2$ is real
which gives rise the propagating modes. Consequently, some
resonance states are expected to emerge. The resonance states
obey the following resonance condition, $\alpha_2(\theta_1,\xi,E)
w=n\pi$, where $n$ is the resonant order. As seen in
Fig.(\ref{T_theta_xi}a,b), the velocity barrier is transparent
against the propagation of Dirac fermionic waves at the resonance
states. The resonance states emerges at the special values of the
incidence angle, the barrier width and those velocities belonging
to the range of $\xi<1$. To distinguish the propagating from the
evanescent modes, we study conductance as a function of the
barrier width in Fig.(\ref{T_theta_xi}c) for several values of
velocity. For $\xi<1$, conductance has an oscillatory behavior
with the barrier width originating from the propagating modes. on
the other hand, conductance drops sharply to zero for $\xi>1$
which is a trace of the evanescent modes inside the barrier.

\begin{figure}
\centering
\includegraphics[width=7cm]{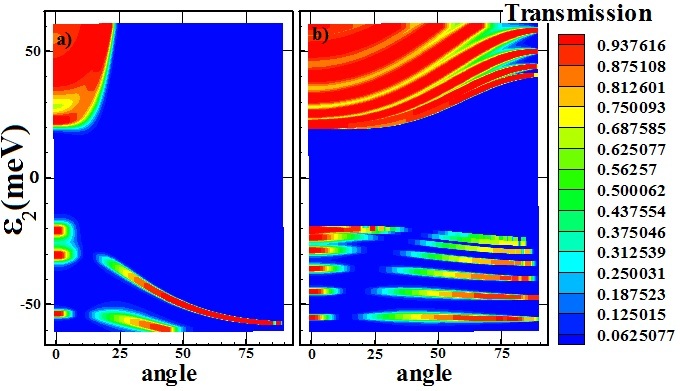}
\includegraphics[width=6cm]{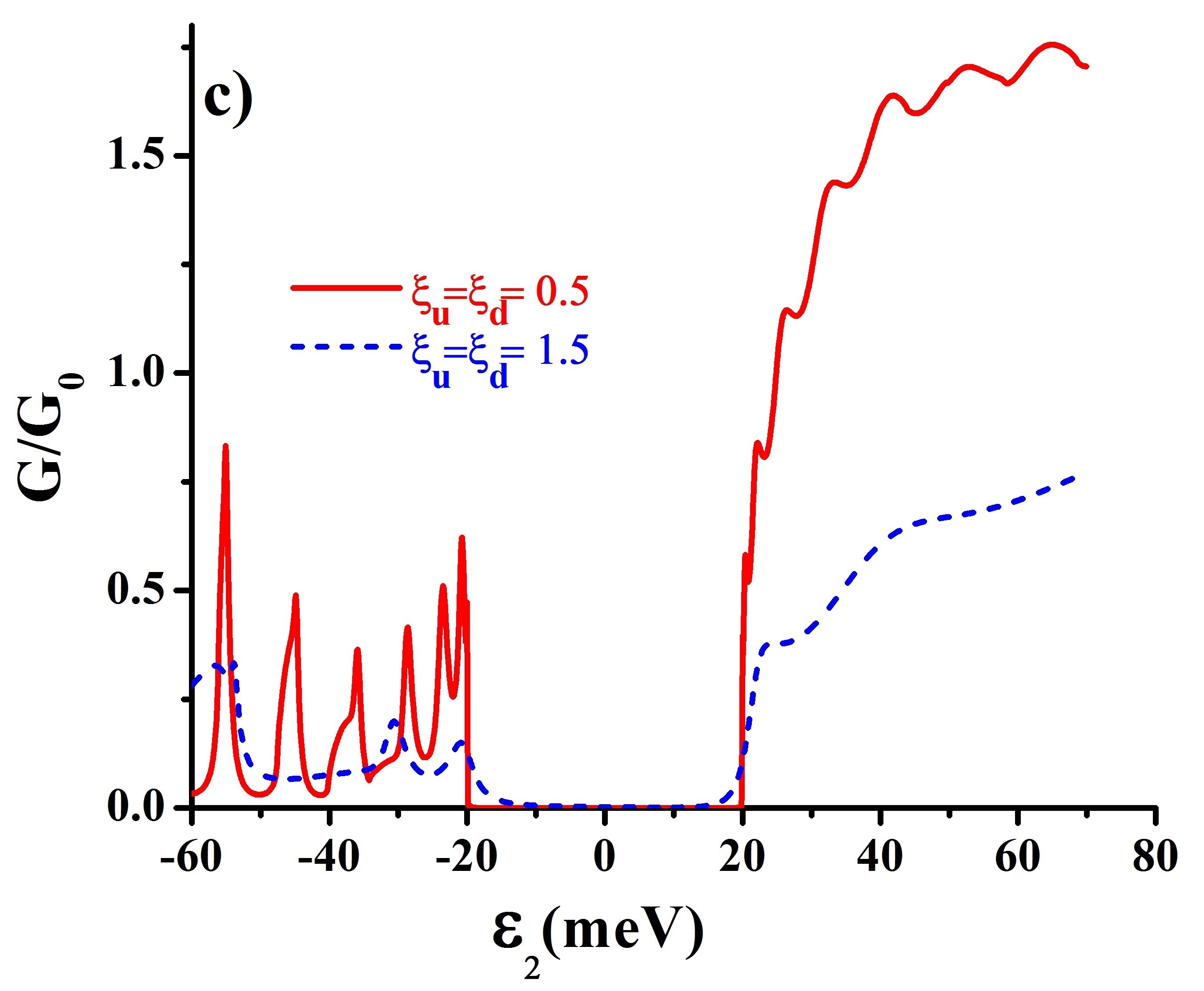}
\caption{a) A 3D contour-plot of transmission probability as a
function of incidence angle $\theta_1$ and energy
$\varepsilon_2=E-V^{'}_0$ for the velocity ratio a)
$\xi_u=\xi_d=1.5>1$ and b) $\xi_u=\xi_d=0.5<1$. c) Conductance in
terms of Fermi energy for the velocity ratio $\xi>1$ and $\xi<1$.
The gate bias is set to $\delta=40 meV$. The gate potential
applying on both layers is set to $V^{'}_0=80 meV$ for inside the
barrier and $V_0=0$ for outside the barrier. The barrier width is
$w=50 nm$. }\label{T_theta_xi_delta}
\end{figure}
\begin{figure}
\centering
\includegraphics[width=7cm]{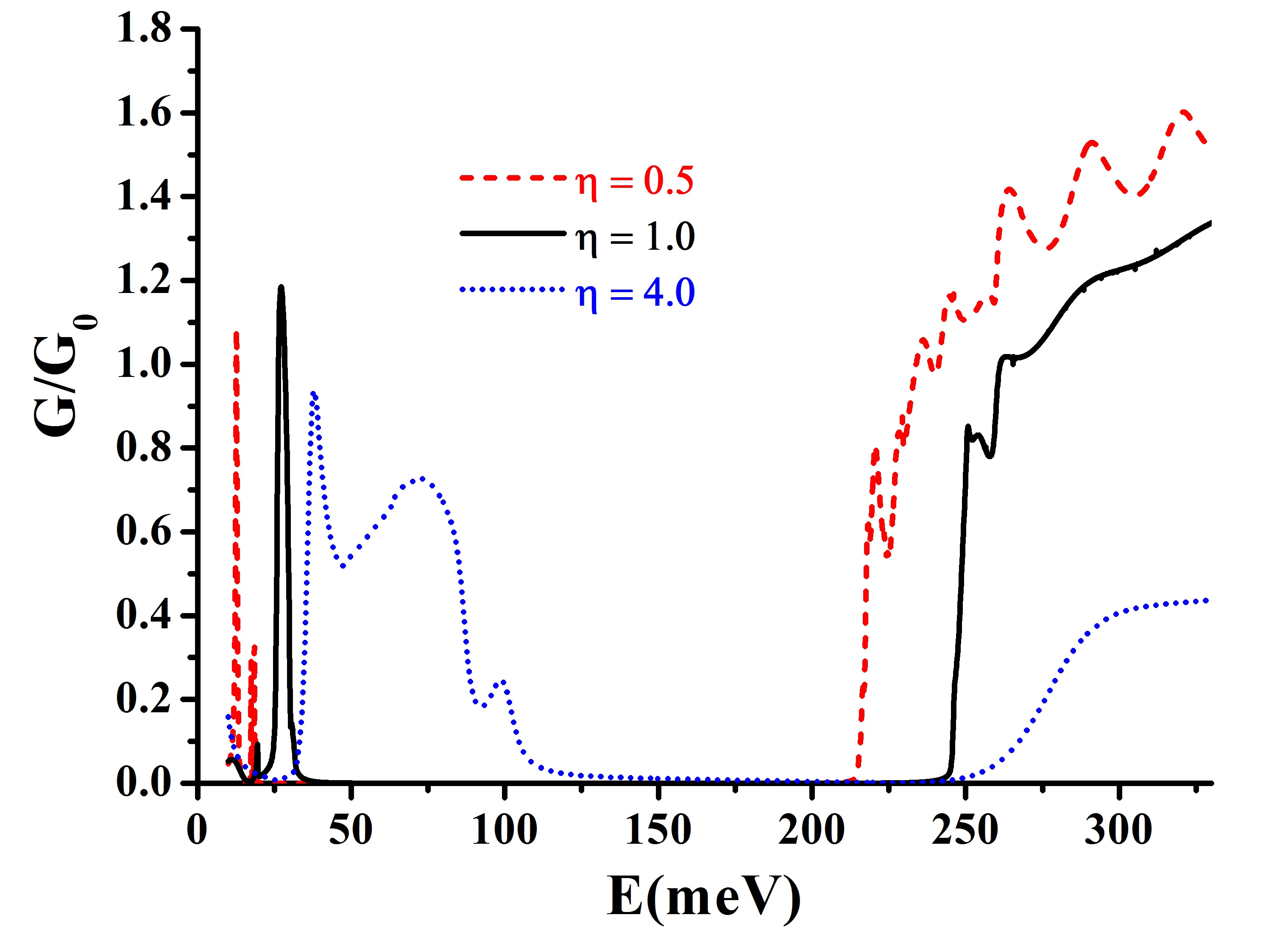}
\caption{ Conductance in terms of Fermi energy for several values
of the velocity ratio $\eta$. Transport gap depends on the
velocity ratio. The gate bias is equal to $\delta=125 meV$. The
gate potential applied on both layers is $V^{'}_0=140 meV$ for
inside the barrier and $V_0=0$ for outside the barrier. The
barrier width is $w=30 nm$. }\label{transport_gap}
\end{figure}

\subsection{Transport across velocity barrier in presence of a gate bias }
{\bf Case ii}: In this case, velocity of carriers changes (still
$\eta=1$) in the barrier region where a perpendicular gate bias is
simultaneously applied ($\delta \neq 0$). To clarify transport
properties of the mentioned system, it is worth to look at the
wave number inside the barrier. The wave number along the x-axis
direction is presented as

\begin{equation}
\alpha _{2}=k\sqrt{\frac{\mu}{\lambda\xi^2}-\sin^{2}\theta_1}
\end{equation}
where
$$\mu=\varepsilon_{2}^{2}+\delta^{2}+\sqrt{{4\varepsilon_{2}^{2}\delta^{2}}-t^{2}(\delta^{2}-\varepsilon_{2}^{2})}$$
$$\lambda=\varepsilon_{1}^{2}+t\varepsilon_{1}$$
Following procedure of the previous section gives a critical
incidence angle as $\theta_{cr.}(E)=
\arcsin(\sqrt{\dfrac{\mu}{\lambda}}\dfrac{1}{\xi})$. Note that in
this formula, the critical angle depends on the Fermi energy.
Therefore, in the presence of electrostatic gate potentials, this
set-up is not proposed as an appropriate material for designing
waveguide. However, a definition of the critical angle for such
system is useful to interpret the behavior of transmission.

A $3D$ contour plot of transmission in terms of the incidence
angle and energy is indicated in Fig.\ref{T_theta_xi_delta} for
two velocity values: a) $\xi=1.5>1$ and b) $\xi=0.5<1$. For
$\xi=1.5$ and for energies greater than the gap
$\varepsilon_2>0$, the existence of a critical angle is clearly
demonstrated in Fig.(\ref{T_theta_xi_delta}a) which shows sharp
dropping of transmission to zero. However, such a sharp critical
angle is absent for $\xi=0.5$. As a conclusion, in addition to the
parameters $\mu$ and $\lambda$ which are energy dependence, the
velocity $\xi$ still can play an important role to tune
transporting modes. To manifest such a property, we study
conductance as a function of energy in
Fig.(\ref{T_theta_xi_delta}c) for $\xi=0.5$ and $1.5$. It is
interesting that conductance for the velocity $\xi=0.5$, is much
greater than its value for $\xi=1.5$. Moreover, conductance has
an oscillatory behavior with the Fermi energy if $\xi<1$. However,
it behaves smoothly with the energy if $\xi>1$.

{\bf Case iii}: In the last case, in addition to the gate biasing
($\delta\neq0$), we modulate velocity in layers not to be equal
to each other $\eta\neq1$. Since the band gap of the barrier
portion depends on the velocity ratio $\eta$, we expect to
manifest this property by concentrating on the transport gap.
Fig.\ref{transport_gap} represents conductance in terms of energy
for several values of the velocity ratio $\eta$. What is novel is
that the transport gap appeared in conductance depends on the
velocity ratio $\eta$. The behavior of the conduction and valence
band edges with the velocity ratio is in good agreement with
those shown in Fig.(\ref{band-gap-u}c).

Referring to Fig.(\ref{k-ph-delta}a), dependence of the band gap
on the velocity ratio is strong when the gate bias is large. So
one can observe that the transport gap depends on the velocity
ratio. The transport gap is remarkable when a thick velocity
barrier is manipulated in the presence of a large gate bias
$\delta$. Maximum band gap emerges at $\eta=1$.

\section{Conclusion}
In the presence of a previously applied gate bias, the electronic
band structure of bilayer graphene is investigated when
quasi-particles have different Fermi velocity in each layer. We
address that the velocity engineering is one of the inevitable
experimental factors which affects the transport gap in the
broken-symmetry BLG.

In absence of any electrostatic potential, only the modulation of
velocity in layers does not cause to open a band gap. In other
words, the chiral symmetry conserves for purely velocity
modulation $\delta=0$ while this symmetry will break when a gate
bias is subsequently applied on BLG. It should be noted that in
the presence of a gate bias $\delta\neq0$, the electron-hole
symmetry preserves whenever the same velocity is modulated in
both layers; $\eta=1$. In addition, the band structure keeps its
'{\it Mexican hat}' shape with a direct band gap. Moreover, the
band gap is independent of velocity value. The maximum value of
the band gap occurs at $\eta=1$. The momentum attributed to the
band gap is inversely proportional to the velocity.

In a generic case, non-equal velocities in two layers
($\eta\neq1$) result in the transition of the direct-to-indirect
band gap. The band gap depends on the velocity ratio $\eta$ and
has a peak at $\eta=1$. Interestingly, the electron-hole symmetry
fails, however the band structure still keeps its '{\it Mexican
hat}' shape. The shift of momentum from the conduction band edge
to the valence band edge is increased with the gate bias.

In the second part, we elaborate a transfer matrix method to
calculate coherent tunneling through a velocity barrier possibly
subjected to a gate potential. In analogous with optics, we
propose a total internal reflection angle $\theta_{cr.}$ so that
transmission becomes sharply negligible for the incidence angles
larger than $\theta_{cr.}$. The transport gap which is induced by
application of the gate bias in the barrier region, depends on the
velocity ratio.

\section{Acknowledgement}

We highly acknowledge R. Asgari for his useful comments during
improvement of this work. One of the authors, (H.C), thanks the
institute for research in fundamental sciences (IPM) and also the
international center for theoretical physics (ICTP) for their
hospitality and support during a visit in which part of this work
was done. We should also thank M. Barbier for his comments in the
four-band tunneling.

\appendix
 \section{The electron-hole asymmetry}
  \label{sec:appendix1}
\begin{figure}
\centering
\includegraphics[width=6cm]{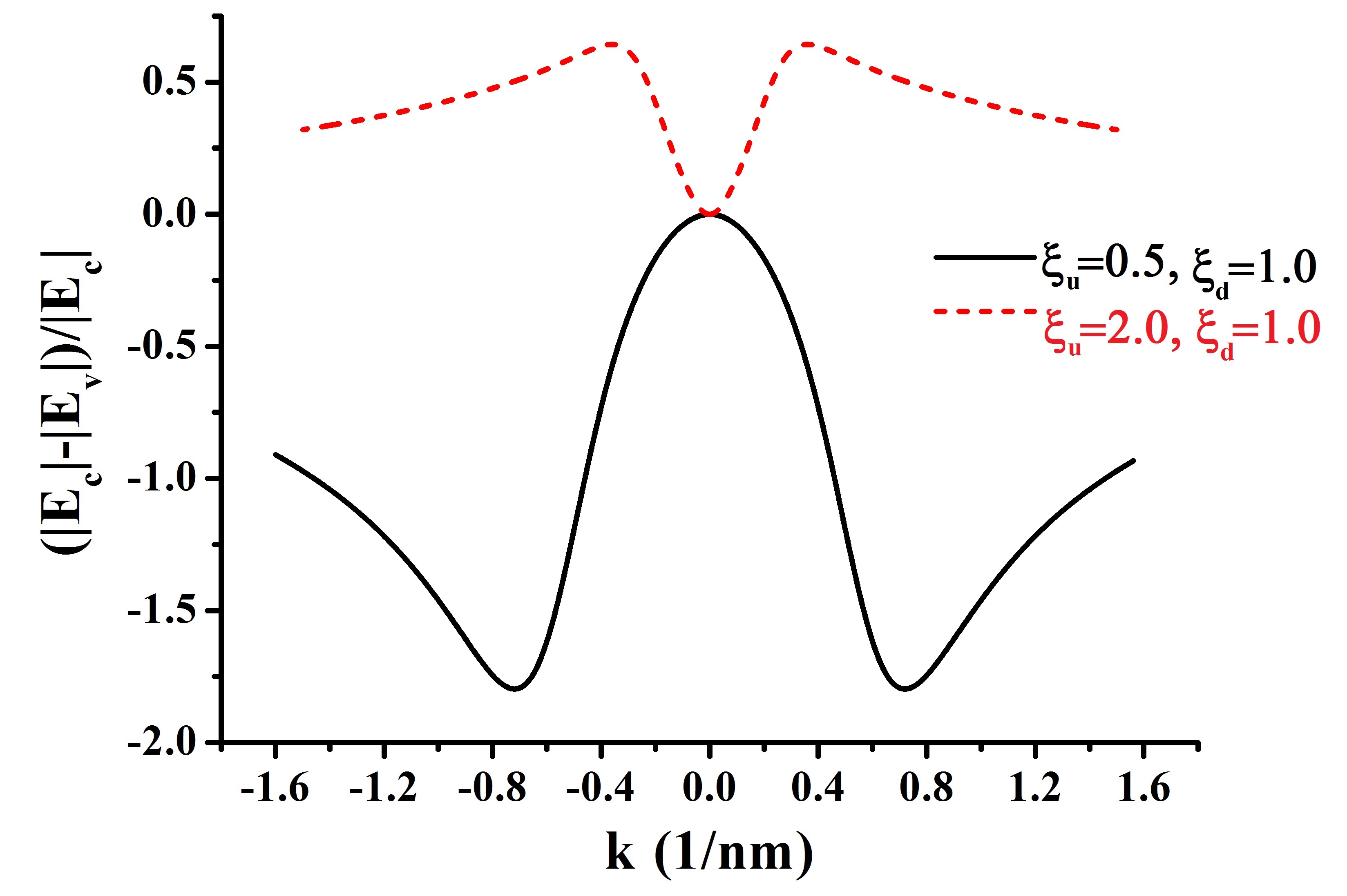}
\caption{ The electron-hole asymmetric factor as a function of
momentum. The gate bias is equal to $\delta=400 meV$. }\label{e-h}
\end{figure}

To measure the electron-hole (e-h) asymmetry, we define the e-h
asymmetric factor as the following; $(\mid E_c \mid-\mid E_v
\mid)/\mid E_c \mid$. For the case of equal velocities modulated
in both layers $\eta=1$, the e-h asymmetric factor is zero for
the studied Hamiltonian shown in Eq.\ref{hamiltonian}. However,
as seen in Fig.\ref{e-h}, this asymmetric factor increases with
the momentum very faster than a linear behavior~\cite{ehasymm}.
This factor reaches to the value of $2$ in the special momentum.
It is interesting that by application of the transformation of
$\eta\rightarrow 1/\eta$, the e-h asymmetric factor behaves as $r
\rightarrow 1/r$.

In addition to the velocity modulation, the e-h asymmetry is also
originated from the inter-layer coupling ($\gamma_4$) between
$A1-A2$ and $B1-B2$ sites~\cite{bilayer-review,ehasymm}. At the
first order approximation, we have not considered such term in
the dominant Hamiltonian shown in Eq.\ref{hamiltonian}. In fact,
the most important terms which affect the main feature of the band
structure are $\gamma_0$ and $\gamma_1=t$. Here, $\gamma_0$ is
the intra-layer hopping between $A1-B1$ and $A2-B2$ sites which
is proportional to $v_F$ in the tight-binding approximation and
$\gamma_1=t$ is the inter-layer coupling between $A2-B1$ sites.
The e-h asymmetric factor caused by parameter $\gamma_4$, behaves
as $4\gamma_4/\gamma_0$~\cite{bilayer-review}. The
well-established values ~\cite{bilayer-review} for the hopping
parameters are equal to $\gamma_4\approx 0.15 eV$ and
$\gamma_0\approx 3 eV$. So the e-h asymmetric factor originating
from $\gamma_4$ is in order of magnitude
$0.1$~\cite{ehasymm,bilayer-review}. As a conclusion, in the
presence of the previously created band gap, the e-h asymmetry
arising from the velocity engineering is a dominant factor in
compared with the e-h asymmetry caused by parameter $\gamma_4$ .

  \section{Wave Function}
  \label{sec:appendix2}
The eigenfunction of four band Hamiltonian of Eq.\ref{hamiltonian}
is defined with the following spinor.
\begin{equation}
\Psi(x) = P(x) A
\end{equation}
where coefficient matrix is written as $$A=\begin{pmatrix}
U_{A_2}&U_{B_2}&D_{B_1}&D_{A_1}
\end{pmatrix}^{\top}$$ and plane wave matrix is presented as
~\cite{cheraghchi-polarization,barbier}
\begin{equation}
P(x)=\begin{pmatrix}
e^{i\alpha_+ x}& e^{-i\alpha_+ x}&e^{i\alpha_- x}&e^{-i\alpha_- x}\\
f_{+}^{+}e^{i\alpha_+ x}&f_{+}^{-} e^{-i\alpha_+ x}&f_{-}^{+}e^{i\alpha_- x}&f_{-}^{-}e^{-i\alpha_- x}\\
s_{+}e^{i\alpha_+ x}& s_{+}e^{-i\alpha_+ x}&s_{-}e^{i\alpha_-  x}&s_{-}e^{-i\alpha_- x}\\
g_{+}^{+}s_{+}e^{i\alpha_+ x}& g_{+}^{-}s_{+}e^{-i\alpha_+ x}&g_{-}^{+}s_{-}e^{i\alpha_- x}&g_{-}^{-}s_{-}e^{-i\alpha_- x}\\
\end{pmatrix}
\end{equation}
\begin{equation}
\begin{array}{c}
 f_{+}^{\pm}=v_{u}\dfrac{\pm\alpha_+-ik_{y}}{\varepsilon-\delta},\,\,\,\,\,
f_{-}^{\pm}=v_{u}\dfrac{\pm\alpha_--ik_{y}}{\varepsilon-\delta}\\
g_{+}^{\pm}=v_{d}\dfrac{\pm\alpha_++ik_{y}}{\varepsilon+\delta}
\,\,\,\,\,\,\,\,\,\,g_{-}^{\pm}=v_{d}\dfrac{\pm\alpha_-+ik_{y}}{\varepsilon+\delta} \\

s_{\pm}=\dfrac{(\varepsilon-\delta)^{2}-v_{u}^{2}[(\alpha_{\pm})^{2}+k_{y}^{2}]
} {t(\varepsilon-\delta) } ,\,\,\,\,\varepsilon_{i}= E-V_{i}
\end{array}
\end{equation}
where $\alpha_+$ and $\alpha_-$ are the wave vectors along the
current direction ($x$) which is defined as
 \beq
\alpha_{\pm} =\sqrt{a(\varepsilon,\eta,\delta)-v_u^2 k_{y}^{2}\pm
\sqrt{a(\varepsilon,\eta,\delta)^{2}-b(\varepsilon,\eta,\delta)}}/v_u
 \label{spectrum2}
\eeq.

If the gate voltage turns on, $a$ and $b$ defined in
Eq.\ref{spectrum} are function of $\varepsilon$ in stead of $E$.

\end{document}